# INVESTIGATION OF ELECTRON BEAM EFFECTS ON L SHELL MO PLASMA PRODUCED BY A COMPACT LC GENERATOR USING PATTERN RECOGNITION

M.F. Yilmaz[a], * , Y. Danisman[b], M. Ozdemir[a], B. Karlık[c], J. Larour[d]

[a]Basic Sciences, Engineering Department, Dammam University, Dammam, KSA

[b]Department of Mathematics, University of Oklahoma, Norman, OK, USA

[c]Independent Researcher, Montreal, QC, CANADA

[d]Laboratoire de Physique des Plasmas (LPP), Ecole Polytechnique, UPMC, CNRS, Palaiseau, France

**ABSTRACT**

In this paper, the effects of electron beam on the X pinch produced spectra of *L*-shell Mo plasma have been investigated by principal component analysis (PCA), and this analysis is compared with that of line ratio diagnostics. Spectral database for PCA extraction was arranged using the non-LTE collisional radiative *L*-shell Mo model. PC vector spectra of L shell Mo including F, Ne, Na and Mg- like transitions were studied to investigate the polarization types of these transitions. PC1 vector spectra of F, Ne, Na and Mg- like transitions resulted in linear polarization of Stokes Q profiles. Besides, PC2 vector spectra showed linear polarization of Stokes U profiles of $2p^5 3s$ of Ne like transitions which were recognized as responsive to the magnetic field (Trabert et al., 2017). 3D representation of PCA coefficients demonstrated that addition of electron beam to the non-LTE model generates the quantized collective clusters which are translations of each other and follow V-shaped cascade trajectories except for the case f = 0.0. The extracted principal coefficients were used as a database for the Artifical Neural Network (ANN) to estimate the plasma electron temperature, density and beam fractions of time integrated spatially resolved *L*-shell Mo *X*-pinch plasma spectrum. PCA based ANN provides advantage in reducing the network topology with a more efficient Backpropagation supervised learning algorithm. The modeled plasma electron temperature is about $T_e \sim 660$ eV and density $n_e = 1 \times 10^{20}$ cm$^{-3}$ in the presence of the fraction of the beams with $f \sim 0.1$ and centered energy of 5 keV.

# I. INTRODUCTION

X pinch discharge experiments on the laboratory or table top scales generate localized high energy density plasmas so called hot spots with the sizes $10^{-4}$ to $10^{-1}$ cm, temperatures 0.1 to 1 keV and electron densities ~ $10^{18}$ to $10^{23}$ cm$^3$. Suprathermal hot electrons with anisotropic velocity distributions are the typical by product of the X pinch produced plasmas. Hot electrons are diagnosed by different experimental and computational methods some of which are *x*-ray emission, electron Breamsstrahlung and *K*α emission, spectropolarimetry and particle-in-cell modelling [2-10]. Polarization spectroscopy is known as one of the main methods to diagnose the state of anisotropy. Besides, collisional radiative model with non-Maxwellian electron distribution was another alternative method to diagnose hot electrons in the emission spectra [11]. On the other hand, Yılmaz et al. showed that the application of principal component analysis (PCA) on the collisional radiative model of resonant transitions of L-shell Cu spectra results in linear Stokes profiles of polarization of Ne-like Cupper spectra. Stokes polarizations sets parameters that can describe the degree and the shape of the polarization completely, and they are found in many applications of astrophysical spectra [12 and 13].

PCA is one of the pattern recognition techniques that is used for reducing the dimension of a dataset of high dimension while keeping great amount of its variability. So, it makes easier to visualize a dataset. It has been used in many fields like robotics, medicine, remote sensing and so on. PCA has also many applications in spectroscopy, especially in unmixing species and decomposing overlapped spectral lines of UV-VIS-NIR spectroscopy, which is critical for spectral fingerprinting [12-13]. Artificial neural network (ANN), in simple terms, is a well-known machine learning algorithm that uses examples to extract rules. ANN is composed of highly interconnected layers that process given examples. In our study, PCA coefficients obtained as a result of PCA analysis and corresponding plasma electron temperature, density and beam fractions from a representative time integrated spatially resolved *L*-shell Mo *X*-pinch plasma spectrum are used as training examples of ANN[15].

In this work, the effects of the electron beam on the non-LTE collisional radiative model of L-shell Mo spectra obtained by PCA have been investigated for a typical X-pinch spectrum (shot XP_633) recorded on a compact low energy device. The plasma electron temperature, density and beam fraction have been extracted using PCA based ANN. The paper is organized as follows. The second chapter summarizes the experiments, and the third chapter studies the effect of the electron beams on the non-LTE K-shell spectra by means of line ratio diagnostics and PCA. The fourth chapter presents the modeling of the experimental data by PCA based ANN, and the conclusions are given in the final section.

## II. EXPERIMENTS

One can find the details of the X pinch experiments of Mo shots in the work of Larour et al 2005. Briefly, the x-ray spectrum of Mo (shot XP_633) was generated by the discharge current of 250 kA with the rise time of 200 ns and the voltage of 40 kV. The 25 mm of Mo wires were placed in the 9 mm of anode-cathode gap in the form of X shape to generate point like plasma in the vicinity of the cross point of the wires. The spectra in the region of 950-1350 eV recorded through two X-ray spectrometers [16 and 17].

A wider spectral region was registered by a convex mica crystal with the curvature radius 2d=1.984 A. The distance from the plasma was 220 mm to achieve effective dispersion of the 25 eV/mm at the 1st order around 1 keV on a cylindrically bent film (R = 28 mm). A narrower spectral region was registered by a flat crystal spectrograph. Specifically, KAP (2d = 2.664 A) and PET (2d = 0.874 A) were used for the Mo experiments. The distanceof the crystal from the plasma was 380 mm and the crystal-film one was 40 mm [16]. The time-integrated spectra were recorded on Kodak Direct Exposure film (DEF). The filtered pinholes(R= 30 μm)on the entrance window of the spectrometer were used to obtain the plasma sizes. The positions of the lines were estimated using the geometry for each shot. Then a consistent set of lines is compared for identification with the database of spectra measured in pulsed hot plasma experiments with X-pinches and Z-pinches [16 and 17].

Fig.1. presents a typical, axially resolved, x-ray spectrum of Mo (shot XP_633) and its corresponding pinhole image. The F- like F1, Ne-like 3C, 3D, 3F and 3G and Na-like Na1 and Na2 transitions are well resolved in the spectrum. The observed $L_\beta$ transition is expected due to suprathermal electron beam effects in plasma [16 and 17]. In figure 2 electrical (voltage, *B*-dot probe signal, current as numerically integrated from *B*-dot) and X ray diode (XRD) records have been illustrated. Fig.2. shows that the main X ray burst happens around 180 ns and magnetic field varies significantly after the main burst.

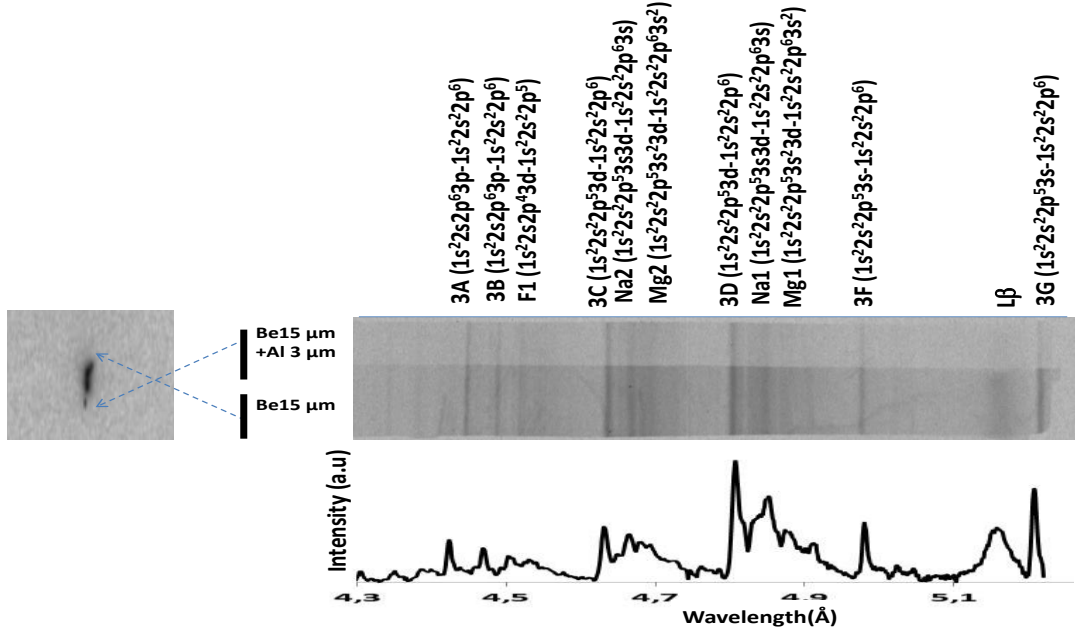

Fig.1. a) Pinhole image b) Time integrated spectra of plasma of Mo XP_633

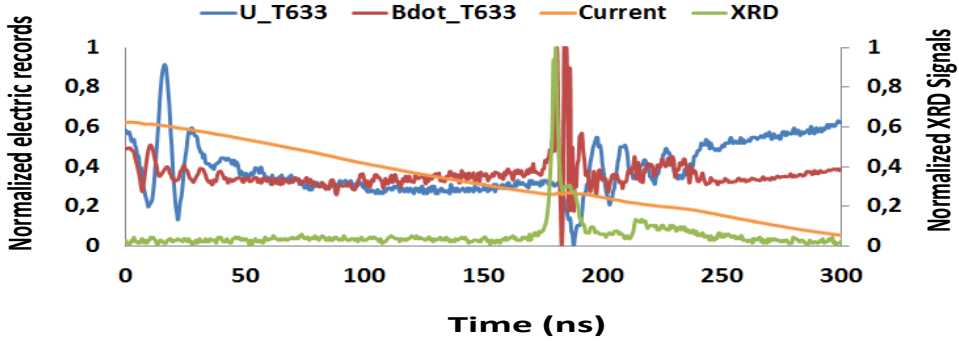

Fig. 2. Close-up of the electrical and photonic records 500 ns around the time of pinching (shot XP_633). The electrical records (voltage, *B*-dot probe signal, current as numerically integrated from *B*-dot). X-ray signal is figured out by the *XRD* signal in volt (right scale).

### III.  ELECTRON BEAM EFFECTS ON THE L-SHELL Mo SPECTRA

#### A. NON LTE MODEL OF MO

The database for modeling of plasma electron density and temperature of Mo (shot XP_633) was generated by using previously developed L-shell non-LTE collisional radiative model. The energy level structures, spontaneous and collisional rates, collisional and photoionization cross-sections calculations were performed using the HULLAC code [18]. The L-shell Mo model includes detailed structures for O-like to Mg-like Mo ions [17]. The model uses hybrid electron distribution function $F(e) = (1-f)*F_{maxwellian} + f*F_{nonmaxwellian}$ to calculate the rates of collisional processes by integrating cross sections over the electron distribution function. In this work, the fraction f of hot electrons was described by a Gaussian distribution centered at the characteristic energy $E_0$= 5 keV. [12,13]. Voigt profiles with the resolution of d = 500 were used to fit line broadening of the experimental spectra [17].

## B. PCA ANALYSIS OF L SHELL MO SYNTHETIC DATABASE

It has been already shown that the ratios Mg1/Na1, F1/Mg1 can be used as electron temperature diagnostics of L Shell Mo plasmas and vice versa [19 and 20]. In this work, we have used the line ratio (Na1+3D)/Mg1 as the plasma electron temperature diagnostics at the moderate electron densities. Fig.3. shows that addition of the beam fraction fixes the this ratio especially Te below 400 eV and the plots tend to have the form of hollow the beam fraction increases.

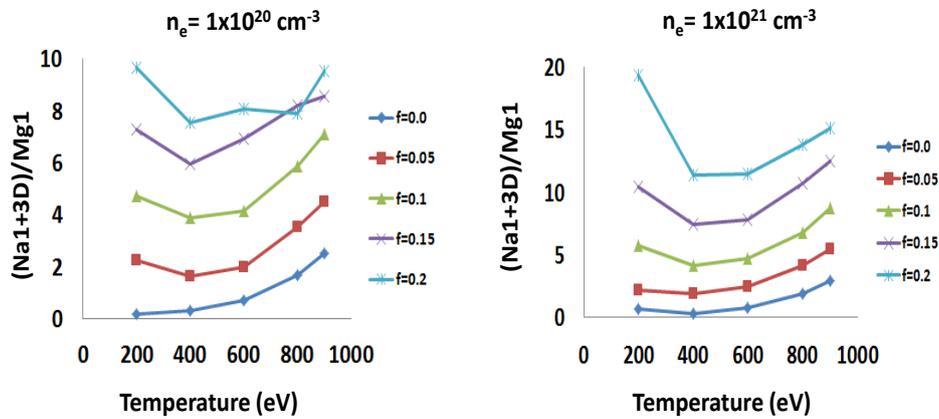

Fig.3. The dependence of the line ratio of (Na1+3D)/Mg1 on plasma electron temperatures.

Yilmaz et al. 2015 described that PCA also can be used as alternative diagnostics instead of line ratio diagnostics for investigating the effects of electron beams on each line dependences and for estimating the plasma parameters[12 and 13]. PCA is a dimension reduction technique of a large databases while retaining most of the information. The main goal of the PCA is diagnosing the hidden structures of the database by linearly transforming the original variables into new uncorrelated variables, called principal components (PC). The Principal components with the big eigenvalues correponds to the maximum varience. Principal components are the eigenvectors of the covariance matrix of the data associate with the largest eigenvalues [21 and 22].

Principal Components ($|PC1>, |PC2>, |PC3>,…$) are used to form the orthonormal basis of the new vector space with smaller dimension. Each original data is projected onto this space and new coordinates are obtained by taking the dot product of the original data and each principal component.

For i=1,2,…,M let $\Gamma_i$ be the vectors in a data set of size N×1. The mean of $\Gamma_i$'s is

$$\mu = \frac{1}{M}\sum_{i=1}^{M} \Gamma_i$$

Now subtract the mean µ from each of $\Gamma_i$ and define

$$\Phi_i = \Gamma_i - \mu$$

The covariance matrix **C** is

$$C = \frac{1}{M}\sum_{i=1}^{M} \Phi_i \Phi_i^t = \frac{1}{M} AA^t$$

where superscript t means transpose and A =[ $\Phi_1, \Phi_2,..., \Phi_M$]. **C** is an N×N symmetric matrix. It is diagnosable and has N nonnegative eigenvalues and eigenvectors. The eigenvector corresponding to the largest eigenvalue is called the first principal component (|PC1, and the second and third largest ones are called the second (|PC2>) and the third (|PC3>) principal component , respectively. If a vector |v> is projected into the space spanned by |PC1>, |PC2>) and |PC3> then we have

$$\text{Proj}_{<|PC1>,|PC2>,|PC3>>}|v> = w_1|PC1> + w_2|PC2> + w_3|PC3>$$

The coefficients $w_1$, $w_2$ and $w_3$ are called the weights of |PC1>, |PC2> and |PC3> in |v> and calculated as

$$w_1 = |v> \cdot (|PC1>)^t \qquad (1)$$

$$w_2 = |v> \cdot (|PC2>)^t, \qquad (2)$$

$$w_3 = |v> \cdot (|PC3>)^t, \qquad (3)$$

where ▪ is the dot product in Euclidean space ($R^N$). Since |PC1>, |PC2> and |PC3> are the most dominant three eigenvectors, the vector v-($w_1$|PC1>+$w_2$|PC2>+$w_2$|PC2>) has less significance and it can be ignored. Therefore, it is enough to work on the three dimensional space spanned by |PC1>, |PC2> and |PC3>. In this work, the PCA is applied to the data obtained for different electron beam fractions separately. Four densities are considered as a training set for a fraction. Each density consists of spectra for the temperatures 200, 220, 240,… 900 eV (36 different temperatures).

One can measure the state of polarization of the light in terms of Stokes parameters. Stokes parameters of I,Q,U and V are used to describe the polarization type of the light. Stokes I represents the unpolarized light, Q and U represent linearly polarized light, and V represents the circulary polarized light[23]. Since PCA can represent the database in a scalar and vectoral manner, one can easily observe the direction change of the spectral lines in vectoral representation of the spectral database [24]. Paletou et al., 2012 showed that PCA is an efficient tool to extract Stokes parameters of the polarized stellar data [25].

Fig. 4. illustrates the vector representations (|PC1> and |PC2>) of L-shell Mo spectra. Mean spectra represent the unpolarized (Stokes I) spectra. |PC1> spectra show that addition of electron beams results in linear polarization of Stokes Q profiles for the considered transitions of F, Ne, Na and Mg-like L Shell Mo. |PC2> spectra which is orthogonal |PC1> spectra show that the 3F and 3G of Ne-like Mo have the direction changes from positive to negative, and Mg2-like Mo has a direction change from negative to positive, which is described as Stokes U. The 3F and 3G of Ne transitions have been already recognized as sensetive to the magnetic fields [1]. Our data in B-dot signals clearly shows a significant variation of the induced magnetic fields after the main x- ray burst, and it was known that hot electron flux supports self generated magnetic fields [26 and 27]. For these reasons, IPC2> vector spectra are expected to represent the the propagation of the photons along the induced magnetic field.

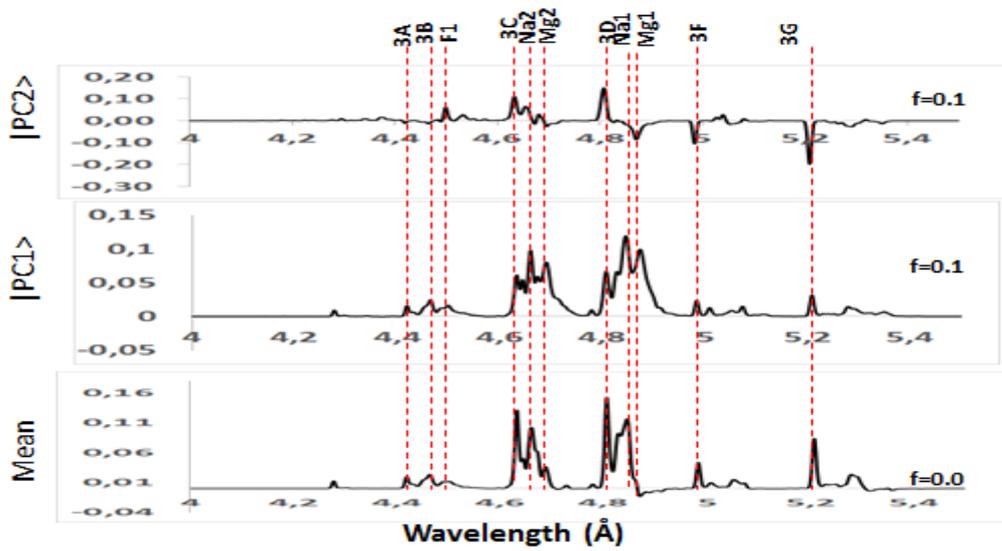

Fig. 4. Mean, IPC1> and IPC2> spectra, a) without beam fraction, f = 0.0 and b) with beam fractions, f = 0.1,

Fig.5. shows the |PC1>, |PC2>, |PC3> coordinates of original data for different fractions at electron density of $n_e = 1 \times 10^{20}$ cm$^{-3}$. In particular, each cluster corresponds to a fraction of electron beams. The clusters of the fractions (except f = 0.0) are the translations of each other, and they form V-shaped cascade trajectory. This shows that addition of electron beams stimulate collective behavior. Such V- shapes are observed in coronal bursts and described by the two stream instability due to collective and hybrid nature of photon and plasmon interactions[24]. Gedik et al, 2016 illustared experimentally periodic V shape like Dirac cones was due to the interaction of photons with the free electrons (plasmons) of the Floquet-Bloch states of topological insulators. Furthermore, these free electrons selectively scatters between Floquet-Bloch and Volkov states [25 and 26].

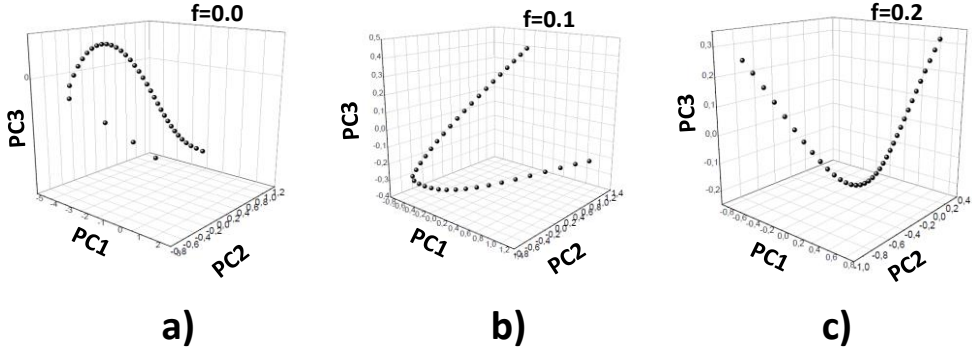

Fig. 5. 3D representation of |PC1>, |PC2> and |PC3> coefficients for different electron beam fractions (f = 0.00, 0.05, 0.10, 0.15 and 0.20) at electron density of $n_e = 1 \times 10^{20}$ cm$^{-3}$.

Fig.6. shows the electron temperature dependence of PC1 coefficients for the considered electron densities. As the fraction of the beam increases, the PC1 coefficients tend to have the shape of the curve for the fractions f=0.05 and f=0.1. Such tendency is in an agreement with the line ratio diagnostics (Na1+3D)/Mg1. However, addition of the beam fraction linearize the PC1 coefficients for f=0.15 and f= 0.2.

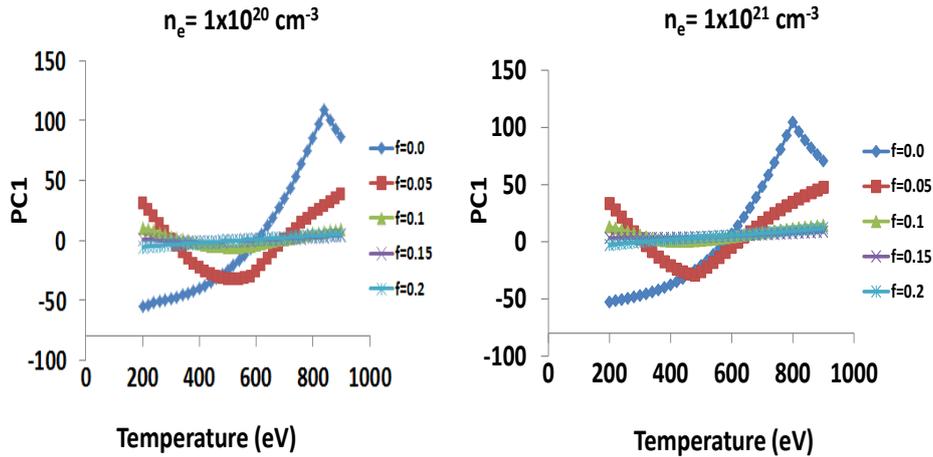

Fig. 6. The correspondence of |PC1> coefficients and electron temperatures at classified electron densities and beam fractions

## IV. PCA BASED ANN MODELING OF MO XP_633

ANN is a great alternative method for classification, prediction and forecasting (nonlinear processing property)[31]. They are powerful tools that can be train given data to perform various tasks such as boundary and feature extraction, information retrieval and many other pattern recognition problems[32 and 33]. The feed-forward neural networks (FFNN) is a suitable structure for nonlinear separable input data. In FFNN model, the neurons are organized in the form of layers. The neurons in a layer get input from the previous layer and feed their output to the next layer as shown in Fig.7. In this type of networks, connections to the neurons in the same or previous layers are not permitted. Learning process in FFNN is back-propagation which requires providing pairs of input and target vectors.

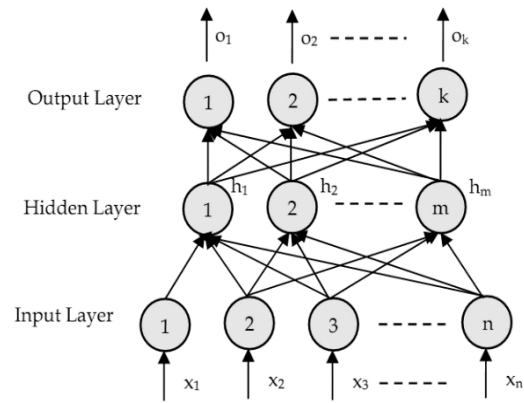

Fig.7. Architecture of a FFNN for classification

In this study, three-layer feed-forward neural network was used for our purpose by using the error back-propagation algorithm. We have employed PCA based neural network to obtain the plasma parameters of the experimental spectra of *X*-Pinch *L*-shell Mo spectra. In the Back-propagation supervised learning algorithm with sigmoid transfer type has been chosen as the activation function. The activation function is used to transform the activation level of a unit (nodes or neuron) into an output signal[34]. The gradient descent with momentum weight and bias learning function are used in back-propagation algorithm. Momentum constant was set to 0.95 while the learning rate was set to 0.01. Mean square error MSE was set to $1{,}5 \times 10^{-5}$, while the number of epoch was selected as 2000.

The principal coefficients of the experimental spectra are computed by taking the dot product of principal components and the difference between this spectrum and the mean of the original 144 spectra. Then, the first principle component of experimental spectra tested by ANN to estimate the plasma electron temperature, and ANN gives Te=659,89 eV ~660 eV and ne=$1 \times 10^{20}$ cm$^{-3}$ and f=0.1 of hot electrons. Fig.8. illustrates the experimental spectrum of XP_633and its modeling by PCA based ANN. It is found that the percentage error between experiment and modeling is 16.5 %.

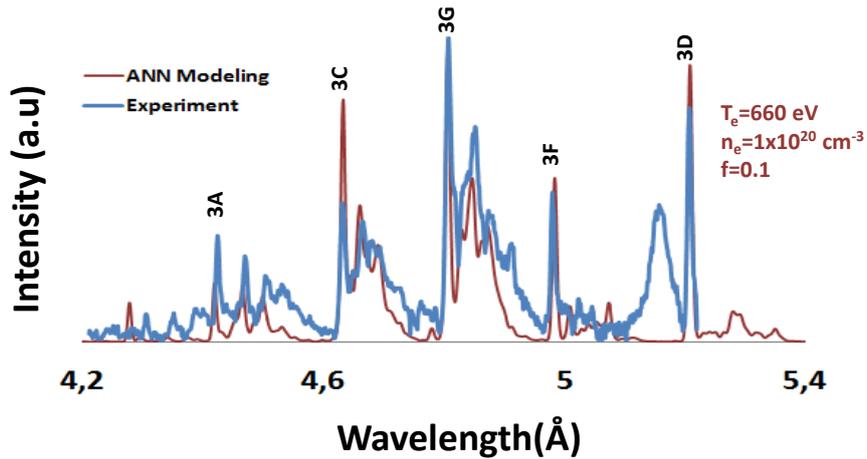

Fig.8. Comparison of experimental spectrum of XP_633 and ANN produced synthetic spectrum (Te=660 eV, ne = $1 \times 10^{20}$ cm$^{-3}$ and f = 0.1).

**CONCLUSION**

PCA can be used for the data classification of non-LTE collisional radiative L-shell Mo model, and each spectrum can be characterized by the dominant PC coefficients. The comparison of PCA with the line ratio diagnostics shows that PCA can be used as an alternative plasma diagnostics of L-shell Mo spectra. The plot of |PC1>, |PC2> and |PC3> coefficients (at electron density of $n_e = 1 \times 10^{20}$ cm$^{-3}$) clearly shows that addition of electron beam on the spectral model generates quantized clusters. F, Ne, Na and Mg-like L Shell Mo vector spectra tend to have linear polarization of Stoke Profiles in the presence of electron beams which have also been observed with the Ne-like Cu L shell spectra as described in our previous works.

**LIST OF REFERENCES**